\documentclass[12pt,pdftex]{article}
\usepackage{html}
\usepackage{epsf}
\usepackage[pdftex]{graphicx}
\usepackage{amssymb}
\usepackage{hyperref}
\begin{document}
\title{MATTERS OF GRAVITY, The newsletter of the APS Division of Gravitational Physics}
\begin{center}
{ \Large {\bf MATTERS OF GRAVITY}}\\ 
\bigskip
\hrule
\medskip
{The newsletter of the Division of Gravitational Physics of the American Physical 
Society}\\
\medskip
{\bf Number 47 \hfill June 2016}
\end{center}
\begin{flushleft}
\tableofcontents
\vfill\eject
\section*{\noindent  Editor\hfill}
David Garfinkle\\
\smallskip
Department of Physics
Oakland University
Rochester, MI 48309\\
Phone: (248) 370-3411\\
Internet: 
\htmladdnormallink{\protect {\tt{garfinkl-at-oakland.edu}}}
{mailto:garfinkl@oakland.edu}\\
WWW: \htmladdnormallink
{\protect {\tt{http://www.oakland.edu/?id=10223\&sid=249\#garfinkle}}}
{http://www.oakland.edu/?id=10223&sid=249\#garfinkle}\\

\section*{\noindent  Associate Editor\hfill}
Greg Comer\\
\smallskip
Department of Physics and Center for Fluids at All Scales,\\
St. Louis University,
St. Louis, MO 63103\\
Phone: (314) 977-8432\\
Internet:
\htmladdnormallink{\protect {\tt{comergl-at-slu.edu}}}
{mailto:comergl@slu.edu}\\
WWW: \htmladdnormallink{\protect {\tt{http://www.slu.edu/colleges/AS/physics/profs/comer.html}}}
{http://www.slu.edu//colleges/AS/physics/profs/comer.html}\\
\bigskip
\hfill ISSN: 1527-3431

\bigskip

DISCLAIMER: The opinions expressed in the articles of this newsletter represent
the views of the authors and are not necessarily the views of APS.
The articles in this newsletter are not peer reviewed.

\begin{rawhtml}
<P>
<BR><HR><P>
\end{rawhtml}
%{\bf \Large Contents:}
\end{flushleft}
\pagebreak
\section*{Editorial}

The next newsletter is due December 2016.  This and all subsequent
issues will be available on the web at
\htmladdnormallink 
{\protect {\tt {https://files.oakland.edu/users/garfinkl/web/mog/}}}
{https://files.oakland.edu/users/garfinkl/web/mog/} 
All issues before number {\bf 28} are available at
\htmladdnormallink {\protect {\tt {http://www.phys.lsu.edu/mog}}}
{http://www.phys.lsu.edu/mog}

Any ideas for topics
that should be covered by the newsletter should be emailed to me, or 
Greg Comer, or
the relevant correspondent.  Any comments/questions/complaints
about the newsletter should be emailed to me.

A hardcopy of the newsletter is distributed free of charge to the
members of the APS Topical Group on Gravitation upon request (the
default distribution form is via the web) to the secretary of the
Topical Group.  It is considered a lack of etiquette to ask me to mail
you hard copies of the newsletter unless you have exhausted all your
resources to get your copy otherwise.

\hfill David Garfinkle 

\bigbreak

\vspace{-0.8cm}
\parskip=0pt
\section*{Correspondents of Matters of Gravity}
\begin{itemize}
\setlength{\itemsep}{-5pt}
\setlength{\parsep}{0pt}
\item Daniel Holz: Relativistic Astrophysics,
\item Bei-Lok Hu: Quantum Cosmology and Related Topics
\item Veronika Hubeny: String Theory
\item Pedro Marronetti: News from NSF
\item Luis Lehner: Numerical Relativity
\item Jim Isenberg: Mathematical Relativity
\item Katherine Freese: Cosmology
\item Lee Smolin: Quantum Gravity
\item Cliff Will: Confrontation of Theory with Experiment
\item Peter Bender: Space Experiments
\item Jens Gundlach: Laboratory Experiments
\item Warren Johnson: Resonant Mass Gravitational Wave Detectors
\item David Shoemaker: LIGO Project
\item Stan Whitcomb: Gravitational Wave detection
\item Peter Saulson and Jorge Pullin: former editors, correspondents at large.
\end{itemize}
\section*{Division of Gravitational Physics (DGRAV) Authorities}
Chair: Laura Cadonati; Chair-Elect: 
Peter Shawhan; Vice-Chair: Emanuele Berti. 
Secretary-Treasurer: Thomas Baumgarte; Past Chair:  Deirdre Shoemaker;
Members-at-large:
Steven Drasco, Tiffany Summerscales, Duncan Brown, Michele Vallisneri, Kelly Holley-Bockelmann, Leo Stein.
Student Members: Megan Jones, Jessica McIver.
\parskip=10pt

\vfill\eject

\section*{\centerline
{DGRAV}}
\addtocontents{toc}{\protect\medskip}
\addtocontents{toc}{\bf DGRAV News:}
\addcontentsline{toc}{subsubsection}{
\it  DGRAV, by Deirdre Shoemaker}
\parskip=3pt
\begin{center}
Deirdre Shoemaker, Georgia Institute of Technology
\htmladdnormallink{deirdre.shoemaker-at-physics.gatech.edu}
{mailto:deirdre.shoemaker@physics.gatech.edu}
\end{center}

We did it!  After years of trying, the APS Topical group on Gravitation has become the APS Division of Gravitational Physics (DGRAV), as announced at the 2016 GGR Business meeting in Salt Lake City.   What did it take to make it to division?  What does it mean to you?  

As a topical group, we needed to achieve 3\% of APS’ total membership in two consecutive years. We achieved our first 3.07\% in January 2015 and then 3.08\% in January 2016.  Once we reached this milestone, we petitioned the APS Council to become a division.  We were the first group in 17 years to petition for division status.    

What's next?  We need you, the DGRAV membership, to approve our new Division Bylaws.  They have mainly changed to reflect the new name of our unit and to add an elected official, the DGRAV Councilor.  As quoted from the bylaws, ``The Division Councilor shall serve as liaison between the Council of the Society and the Executive Committee of the Division. Following each Council meeting, the Division Councilor shall report to the Chair and the Secretary-Treasurer regarding Council actions that affect the status and operations of the Division. Reports shall be made to the entire Executive Committee during their regularly scheduled meetings.''  The term of the Councilor is four years, beginning on January of the year following the election.  The Councilor may not serve more than two consecutive terms.  

Our new bylaws have one additional change to include the position of webmaster. Many of you may not be aware that this Newsletter is our official newsletter for the APS.   Most of the units in APS have APS handle their newsletters, but Matters of Gravity actually predates the GGR!   In the same spirit of the position of Editor of the Newsletter standing outside of the executive committee of DGRAV, we have added a webmaster. The webmaster has the duty of managing our website, \htmladdnormallink 
{\protect {\tt {http://dgrav.org}}}
{http://dgrav.org}, and our social media presence.

Beverly Berger formed and chaired GGR in 1995.   Let’s give her and all the others (yes including any of your family and friends that joined GGR!) a virtual round of applause for making this possible.

\section*{\centerline
{we hear that \dots}}
\addtocontents{toc}{\protect\medskip}
\addcontentsline{toc}{subsubsection}{
\it we hear that \dots , by David Garfinkle}
\parskip=3pt
\begin{center}
David Garfinkle, Oakland University
\htmladdnormallink{garfinkl-at-oakland.edu}
{mailto:garfinkl@oakland.edu}
\end{center}

Ronald Drever, Kip Thorne, and Rainer Weiss were awarded the Shaw Prize in Astronomy and the Kavli Prize in Astrophysics.

Emanuele Berti was elected Vice Chair of DGRAV; Kelly Holley-Bockelmann and Leo Stein were elected members at large of the Executive Committee of DGRAV. Megan Jones was elected Student Representative of DGRAV. 

Gregory Adkins has been awarded the APS Prize for a Faculty Member for Research in an Undergraduate Institution.  Raymond Beausoleil has been awarded the APS Distinguished Lectureship Award on the Applications of Physics.  

Douglas Finkbeiner, Shane Larson, Pierre Michel, Dwight Neuenschwander, Scott Ransom, Stephan Schlamminger, and Rodger Thompson have been elected APS Fellows.

Hearty Congratulations!

\section*{\centerline
{GW150914 - A Watershed Event for Gravity}}
\addtocontents{toc}{\protect\medskip}
\addtocontents{toc}{\bf Research Briefs:}
\addcontentsline{toc}{subsubsection}{
\it  GW150914 , by Gabriela Gonz\'alez and David Reitze}
\parskip=3pt
\begin{center}
Gabriela Gonz\'alez, Louisiana State University
\htmladdnormallink{gonzalez-at-lsu.edu}
{mailto:gonzalez@lsu.edu}
\end{center}
\begin{center}
David Reitze, LIGO Laboratory and Caltech
\htmladdnormallink{reitze-at-ligo.caltech.edu}
{mailto:reitze@ligo.caltech.edu}
\end{center}

\section*{Introduction}

Those of us who work on LIGO will forever remember exactly where we were on September 14, 2015 when we first learned of a nearly simultaneous `trigger' recorded on the LIGO Hanford and Livingston detectors. That trigger would eventually become GW150914, the first gravitational wave ever recorded. Perhaps equally momentous, it would also reveal the first ever observation of a binary black hole system in the universe colliding to merge and form a new black hole. 

Almost exactly one hundred years after gravitational waves were first theorized by Einstein, the discovery by the LIGO Scientific Collaboration and Virgo Collaboration marked the culmination of a scientific quest that began over 50 years ago. Over that time period, this pursuit has brought together well over a thousand researchers worldwide to develop the new science of gravitational wave physics and astronomy.  And like most endeavors of this historical magnitude, the path had a few twists and turns along the way.  

\section*{History}

Prior to 1960, no one seriously contemplated developing detectors for gravitational waves because, quite simply, no one seriously believed that gravitational waves could ever be detected.  That changed when Joseph Weber began using large cylindrical aluminum bars to search for gravitational waves.  While his claims of gravitational wave detections in the 1960s and 70s ultimately proved to be incorrect (resulting in some acrimonious scientific debates), Weber's experimental efforts led Michael Gertsenshtein and Vladislav Pustovoit and, independently Weber himself and also Rainer Weiss to propose using laser interferometers as detectors. 

Gravitational-wave interferometer research programs sprung up in the 1970s at MIT (Weiss), the University of Glasgow (Ron Drever and Jim Hough), and the Max Planck Institute in Garching, Germany (Hans Billing).  Independently, Kip Thorne (Caltech) began a research group at Caltech focusing on gravitational wave theory, and began collaborating with Vladimir Braginsky (Moscow State University) on some of the more intriguing quantum aspects of suspended-mirror gravitational-wave interferometers. This ultimately blossomed into an experimental effort at Caltech, led by Drever (who had moved from Glasgow) and Stan Whitcomb.  Each of these groups began tackling the challenge of building and understanding the complex subtleties of operating ultrasensitive suspended mirror interferometers. 

The period between 1984 and 1992 saw both innovative advances in interferometer designs and the formulation of a joint Caltech-MIT collaboration to design and build two kilometer-scale gravitational-wave observatories.  However, funding for LIGO didn't come about quickly or easily.  When LIGO was first proposed as a large-scale project, it was met with great resistance. It was deemed too risky and too expensive; the chance of failure was too high, and the scientific payoff too low relative to more established types of astronomy to justify the expenditure.  Nonetheless, the US National Science Foundation recognized both the huge scientific potential in gravitational wave physics and astronomy and the cutting edge technology that could result from designing and building a gravitational wave detector.  The NSF took a huge risk in funding LIGO, but it was a measured risk. The technological leap from the prototypes to the advanced detectors was deemed too great to be carried out in a single step.  A two stage approach was adopted in which an initial set of interferometers (Initial LIGO) would be built with a sensitivity where gravitational waves might be detected (but more likely not), followed by the construction of a second set of interferometers (Advanced LIGO) that would have a high probability of detection. 

In 1991 the US Congress appropriated LIGO's first year of funding.  In 1992 Hanford, Washington and Livingston, Louisiana were chosen as the sites for LIGO's interferometers, and a cooperative agreement for the management of LIGO was signed between NSF and Caltech.  In 1994 Barry Barish (Caltech) was appointed LIGO Director and oversaw LIGO's construction phase as well as the installation and commissioning of LIGO's initial interferometers. In 1997, the LIGO Scientific Collaboration was created to organize and coordinate LIGO's technical and scientific research and data analysis, and for expanding LIGO to include scientists from institutions beyond Caltech and MIT.  Initial LIGO was operated from 2002 through 2010, producing over 100 papers and quite a few interesting upper limits on gravitational wave emissions from compact binary systems, pulsars, and even the primordial universe.  

The initial LIGO interferometers were decommissioned in 2010 to make way for the Advanced LIGO interferometers. Designed to be ten times more sensitive to gravitational wave strains, Advanced LIGO is completely new - every component and subsystem has been re-designed and rebuilt to achieve the sensitivity goal. Following an installation period lasting four years, the Advanced LIGO interferometers were completed in 2014 and commissioned until September 2015 when the inaugural observing run `O1' began. 

\section*{The Advanced LIGO Interferometers}

The Advanced LIGO interferometers are the most sensitive scientific instruments ever conceived and built. Consisting of two identical 4 km arm length interferometers in Hanford, WA and Livingston, LA, they are a technological tour-de-force, bringing together a wide array of technologies that have redefined the state-of-the-art throughout almost every facet of their design - the world’s most stable high power lasers, the most precisely figured and coated 'test mass' mirrors, the most sophisticated low frequency seismic isolation and mirror suspension systems, one of the world’s largest high vacuum systems, and gluing it all together, hundreds of feedback control loops that are capable of sensing and maintaining the 4 km length of the interferometers to almost ${10}^{-19}$ m in their most sensitive bands.

Owing to its cutting edge nature, the successful construction and commissioning of Advanced LIGO required solving a very large number of problems - related to the handling of high optical power, angstrom-level polishing of massive optical components, development of strong but exquisitely delicate silica fibers for suspending the 40 kg test masses (three suspensions experienced fiber breakage during the installation phase despite stringent protocols), and precision control engineering in a high-vacuum environment.  

In the end, everything came together!  The superb LIGO commissioning team made rapid progress, and by September 2015, the Livingston and Hanford interferometers were achieving sensitivities four times better than achieved by initial LIGO.  During the first week of September, the decision was made to officially begin the run on September 18, 2015. The interferometers were in an `Engineering Run' phase, and were taking science quality data since mid-August. 

This turned out to be very fortunate.

\section*{The Discovery}

During `Engineering Run 8', which began on August 17, operations were conducted 24 hours a day and seven days a week. Most of those hours were dedicated to tests for calibrating the detector, tuning the injection methods for simulating gravitational waves in the detector, setting up automated alerts for possible gravitational wave detections, measuring the effects of induced environmental noise in the detector, and many more tasks that needed to be ready before the official first `O1' run.   During the times that the two LIGO detectors were running unperturbed by these tests, online algorithms were constantly running to search for gravitational waves and test their performance against the new Advanced LIGO data. When coincident ``triggers'' produced by these methods exceeded a (low) significance threshold, automated database entries were produced with a lot of numbers and plots - there had been several entries with injection tests, as well as weak coincidences that were not statistically significant. 

On September 14, at 5:51am US Eastern time, a program looking for short transients registered a very significant coincidence. Attentive scientists in Europe and early risers in the US noticed the trigger, and produced a time-frequency plot which showed what is expected from a binary coalescence - but a very short duration one, which would ordinarily correspond to two merging black holes. At first glance, this looked much more like an injection than an astrophysical signal - there was no injection label associated with the trigger, but could it still be an injection? Perhaps a blind injection? Many emails and phone calls later, it was clear that it was not an injection - it was a real coincidence! The next question that was asked - Could it be an instrumental artifact? - was also eventually ruled out. On September 14, many decisions were taken to start the ``discovery process'': the instrument configuration was frozen as much as possible to make sure there was not an instrumental source of transients, coincident or not, that looked like binary coalescences. After a few days, a minimal set of tests was finished, and the instrument was put into ``normal'' observation mode. 

\begin{figure}[ht!]
\centering
\includegraphics[width=0.5\textwidth]{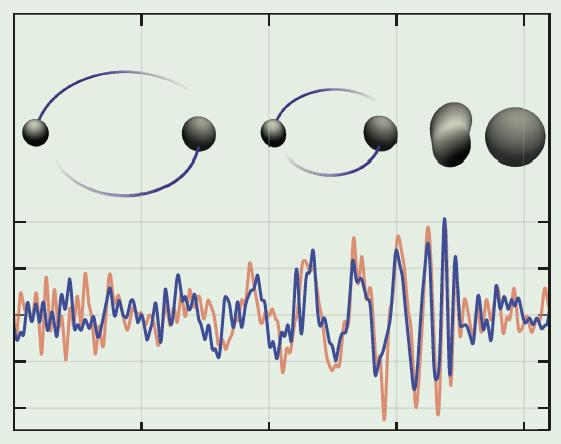}
\caption{Figure on cover of Physical Review Letters, Feb 12, 2016, showing the data in the LIGO Hanford (blue) and Livingston (red) detectors due to the coalescence of two black holes.}
\end{figure}

We estimated we needed at least 15 days of two-interferometer coincident data with no similar transients to bound the significance at the magical 5-sigma threshold. By the end of October, we had collected a sufficient amount of data, and separate `offline' analyses using matched-filtering to look for binary coalescences were ready to `open the box' to answer the question - did the significance hold up?  It did!  A few champagne corks popped, but we were not done yet - long periods over the holidays were still needed to finish the review of the methods used to find the event, review the data calibration and its errors, estimate the parameters of the system, write a paper with the observations and other papers with the details, and wait for external peer review of the main result. While the LSC and Virgo had approved in early September 2015 (!) a ``detection procedure'' to validate the first detection, publish and announce it, very few if any of us thought that we'd have to apply it so soon.  But we were very happy to do so!

Readers of this newsletter have probably already read the articles containing the details: on February 11, the LIGO Scientific and Virgo Collaborations proudly announced the discovery of two black holes, 29 and 36 solar masses respectively, that had merged into a single black hole more than a billion years ago, producing ripples of space time that passed through Earth and are still traveling through the Universe, after leaving a small but very detectable signal first in the LIGO Livingston detector, and seven milliseconds later in the LIGO Hanford detector. The signals were consistent with Einstein's theory of General Relativity, and using his theory we could derive bounds for many parameters for the system: the initial and final masses, spins, distance, inclination, and (very rough) sky localization. It was a single detection, but contained lots of information!  All of the papers produced by the LIGO and Virgo collaborations on GW150914 are available at \htmladdnormallink {\protect {\tt {http://papers.ligo.org}}}
{http://papers.ligo.org}

\section*{The Future is bright!}

The news of the first gravitational wave detection made it to first page of most of the major newspapers worldwide, resulting in tremendous interest by both the scientific and general public. But this is just the beginning: LIGO and Virgo are still finishing the analysis of the rest of the data - taken until January 12 - when the first observational run finished. The analysis of the remaining O1 data should give us more information about the rates of black hole mergers that can be measured. 

The detectors are designed to be about three times more sensitive than the O1 run, and scientists are hard at work on improving the instruments, and making them more robust. A second observational run is planned for the fall of 2016, lasting about six months - we expect more detections, and are now prepared to become a real Observatory. Soon, the Virgo detector will finish installation and commissioning, and join the network to provide much better sky localization of the sources. As the LIGO and Virgo detectors improve both sensitivity and duty cycle, we will see many black hole mergers - and likely other sources, including binary neutron star or neutron star-black hole mergers, a stochastic background of those mergers, possibly close rotating neutron stars in our galaxy - and perhaps more excitingly, gravitational waves of unknown origin. Farther into the future, but on the horizon, detectors in Japan and India will join the network of ground-based interferometers.  Moreover, the success of the LISA Pathfinder mission bodes well for a future gravitational-wave detector in space.

The future is gravity-bright!

\section*{\centerline
{Remembering Felix Pirani}}
\addtocontents{toc}{\protect\medskip}
\addtocontents{toc}{\bf Obituaries:}
\addcontentsline{toc}{subsubsection}{
\it  Remembering Felix Pirani, by Stanley Deser}
\parskip=3pt
\begin{center}
Stanley Deser, Brandeis University
\htmladdnormallink{deser-at-brandeis.edu}
{mailto:deser@brandeis.edu}
\end{center}

Felix Pirani, who died at age 87 on the last day of 2015, was one of the leaders in the postwar renaissance of General Relativity. When I first met him, at the famous ``GR-0''
1955 Bern conference, we constituted half of the younger generation there (and two of those were mere tourists). A prodigy, Felix entered UBC at 14, writing his first paper as an undergraduate. He learned Relativity at Toronto and Carnegie Tech (where he got his DSc) from Schild.  Synge and Infeld were also important mentors. His thesis was one of the early attempts, and the first serious postwar one, at quantizing GR. A subsequent PhD, with Hermann Bondi at Cambridge, led him into cosmology. This was followed by a year with Synge at DIAS in Dublin, which he devoted to establishing the reality of gravitational radiation, a controversial topic at the time. He then joined Bondi's newly formed group at King's College, London, where he was to remain until his 1983 retirement. In 1967, he became head of the group and produced a steady flow of PhDs (many of whom remained close friends) while obtaining major new results. Two early examples: a brilliant application of the then new Petrov classification to establish, invariantly, the reality of gravitational waves; with Bondi and Ivor Robinson, a pioneering study of wave solutions in GR.  
During the year he spent as visiting professor at Brandeis in the sixties, I had the pleasure of collaborating with him on several papers, thus seeing firsthand his mastery of our field; his 1964 Brandeis Summer Institute lectures on gravitational radiation are still
a classic reference. But his contribution to GR is far greater than the standard references to which his name is associated.  He was one of our giants.
Felix was also a man of many interests, political and artistic; indeed, he started a successful career as a mosaicist upon retirement.﻿

\vfill\eject

\section*{\centerline
{Remembering David Finkelstein}}
\addcontentsline{toc}{subsubsection}{
\it  Remembering David Finkelstein, by Predrag Cvitanovi\'c}
\parskip=3pt
\begin{center}
Predrag Cvitanovi\'c, Georgia Institute of Technology
\htmladdnormallink{predrag.cvitanovic-at-physics.gatech.edu}
{mailto:predrag.cvitanovic@physics.gatech.edu}
\end{center}

Theoretical physicist David Ritz Finkelstein, Professor Emeritus in
the School of Physics of Georgia Tech, died peacefully at home on January
24, 2016. He was born in 1929, in New York City. He went to
Stuyvesant High School in Manhattan, and worked as a page in the NY
Public Library, which gave him access to the stacks where he spent
much time reading. He
started out as an engineering major at City College of New York,
but graduated in 1949 with honors in both physics and mathematics,
and the CCNY Physics Medal. 

In 1953 he received a PhD in Physics from MIT (advisor: Felix Villars), with a
thesis entitled ``Non-linear meson theory of nuclear forces.''  From 1953
to 1960 he worked at Stevens Institute of Technology. He was Young
Men's Philanthropic League Professor of Physics at Yeshiva
University from 1959 to 1976.

In Sidney Coleman's words, he ``was a brilliant scientist with a
passion for long shots. This meant that nine times out of ten he
devoted his talents to ideas that do not pay off, but one time out
of ten, they do pay off. When this happened, David's work was found
to be of a great significance, extraordinary penetration, and ten
years ahead of everyone else's, as was the case when topological
conservation laws entered the mainstream of quantum field theory.''

In a 1955 paper Finkelstein addressed the question of whether
``anomalous'' spin 1/2 had been overlooked for the gravitational field.
His discovery of the topological origin of such anomalous spins and
a speculation that maybe all physical variables are topological in
origin was the thread that led him to kinks and the unidirectional
membrane in the 50s and 60s, as well as to anyons in the 80s,
antecedents of anomalous quantum numbers in the fractional Hall
effect and in high temperature superconductivity.

Finkelstein was the first to understand the nature of the black
hole horizon. In 1958 he (age 29) described what is now known as a
black hole (``unidirectional membrane''). The paper influenced
Landau, Penrose and eventually Wheeler, and was instrumental in
bringing general relativity into mainstream physics and today's
vibrant black-hole research. But it took time. At the 1962
Relativistic Theories of Gravitation annual conference Finkelstein
reported on his Schwarzschild black hole to an audience of three
who turned out to be janitorial staff.

In 1957, following a seminar Finkelstein gave in London, he met
Roger Penrose, then a graduate student from Cambridge. The seminar,
on extending Schwarzschild's metric, both into the past and into
the future across null horizons-a basic ingredient of the current
understanding of black holes---had been a revelation to Penrose.
After the seminar Penrose explained to Finkelstein his
spin-networks, and the two men exchanged their research subjects,
for ever after. Finkelstein's extension of the Schwarzschild metric
provided Penrose with an opening into general relativity, the
subject which has animated his research ever since. Finkelstein
picked up in return on the combinatorial aspects of quantum spin as a
possible route to delving more deeply into the quantum nature of
reality and took such ideas to greater lengths than anyone else.

Another colleague whose career was shaped by Finkelstein's insights
is Lenny Susskind. In 1967 he told Susskind: ``Forget perturbation
theory. Black holes are the key.'' He explained--this was before
Bekenstein--that the information in a region of space could not be
as rich as the volume because most states would collapse to form a
black hole. Finkelstein understood the holographic principle long
before 't Hooft and Susskind.

While his ``unidirectional membrane'' is today considered his key
contribution to physics, for Finkelstein that calculation was only
an exercise, an illustration for his overarching program to bring
topology into quantum physics. He was the first to discover
``kinks'', topological charges and topological spin-statistics
theorems, with Misner (1959) and Rubenstein (1962). Until the work
of Finkelstein and Rubinstein on topology in quantum field theory,
quantum field theory meant Feynman diagrams. He was arguably the
first to understand the role of quantum vacua, and his papers were
among the earliest on solitons in quantum theories, leading to
instantons, Higgs particles, etc... It took quite a few years for
the rest of the world to catch up. The 1962-63 papers with Jauch,
Schiminovich and Speiser were the first to formulate a unified
SU(2) gauge theory of massive vector bosons and light, introducing
a type of ``Higgs mechanism'' before Higgs, and a type of electroweak unification
before Glashow, Salam and Weinberg. However, by the late 1960s the
limitations of the quaternionic formulation led him to shelve the
whole quaternionic project.

In 1976 Finkelstein became the chairman of the Belfer School of Yeshiva
University Physics Department, and in 1978 its dean of Natural
Sciences and Mathematics. On the basis of this administrative
experience, he was appointed Georgia Institute of Technology
Director of the School of Physics.  Nobody understood his job colloquium, 
but all were impressed by recommendation
letters the likes of which Georgia Tech had never seen. No letter mentioned
the fact that his administrative jobs at Yeshiva were ceremonial,
to finalize the closing of the already shuttered graduate sciences
program. The first thing he
did upon arrival was to inform the departmental secretary (in those
halcyon days the department was run by a single secretary) that he
was going to Majorca. So for a month the Chair could not be
reached. But when, by the midyear, he failed to submit a budget, he
was deposed by senior faculty, and replaced by an acting chair.
Finkelstein went into a funk for two weeks, and emerged a changed
man, having recognized that his failure as administrator had
given him more time for research.

He was charismatic and involved a number of dedicated students in
his efforts to quantize geometry. If he did not show up at work,
his students would go to his house, where he met them in his
bathrobe. He was blissfully useless for any faculty committee task. In
Finkelstein's own words: ``My committee services within Georgia Tech
have not been onerous; I do not look this gift horse in the mouth,
but serve as requested by the School of Physics.'' He taught
whatever he wanted to teach, so he had to be deposed for the second
time, this time by an uprising of undergraduates who found
themselves taught quantum logic instead of the introduction to
quantum mechanics that they had signed up for. 

In 2003 Finkelstein noticed in a lecture that between glancing at a
formula in his notes, and writing it on the blackboard, he had
forgotten it. So he went to the Chair, and arranged for his
retirement. Soon enough he found out that it was statins that were
making him stupid - once he went off statins, his sharp intellect
came back. In retrospect, he found this misinformed decision was
one of the best of his life - now he could devote himself fully to
his research.

A few days before his death from idiopathic pulmonary fibrosis he
had his laptop in his bed, and was still working. David was a man
who truly loved life. He is survived by his wife Shlomit Ritz
Finkelstein, his children Daniel Finkelstein, Beth Bosworth, Eve
Finkelstein, and Aria Ritz Finkelstein, his five grandchildren and
two great-granddaughters.

\vfill\eject

\section*{\centerline
{Steve, the Physicist}}
\addcontentsline{toc}{subsubsection}{
\it  Steve, the Physicist, by Bernard Whiting and Eric Poisson}
\parskip=3pt
\begin{center}
Bernard Whiting, University of Florida
\htmladdnormallink{bernard-at-phys.ufl.edu}
{mailto:bernard@phys.ufl.edu}
\end{center}
\begin{center}
Eric Poisson, University of Guelph
\htmladdnormallink{epoisson-at-uoguelph.ca}
{mailto:epoisson@uoguelph.ca}
\end{center}

As a community, we experienced a most bitter-sweet moment earlier this year when the LIGO announcement of gravitational wave detection came just a few days after we had suffered the sudden and unexpected loss of Steven Detweiler as a friend and colleague on February 8th, 2016.

Gravitational wave physics was the area of research which drove Steve all his working life.  Thus, it was with some irony that we found ourselves grieving at his passing while celebrating his life's interest during the announcement of the LIGO detection.  Steve was not a member of the LIGO Scientific Collaboration that finally made the detection because he wanted the freedom to direct his own research. Nevertheless, he was an avid follower of the LIGO experiment and was fortunate enough to have heard, a week or so before his last early morning run, that a detection had been made.  What would have thrilled Steve more than the announcement itself would have been the news that the waves detected actually came from colliding black holes with an unexpected size, about thirty times the mass of our Sun --- not smaller, and not much larger ({\it i.e.}, not supermassive) either.  The study of black holes, and the gravitational waves they produce, had been a constant theme permeating Steve's scientific research.

Steve initially began his work on black holes in conjunction with
the Nobel Laureate, Subrahmanyan Chandrasekhar.  It had been recently realized that when a black hole is disturbed --- say, by something falling in --- it will actually vibrate, and those vibrations initiate waves which propagate away throughout spacetime.  The interaction between each black hole and its surrounding spacetime is very stiff, so the vibrations and gravitational waves are rapidly damped down and soon die away to nothing at all.  But the frequencies and damping times are unique characteristics of the black hole they come from, and Steve was among the first to calculate and tabulate them for later use.  In fact, that ``ring-down'' was one of the tell-tale features which the LIGO collaboration was so anxious to identify, since its unique character ensures that a black hole origin is absolutely unambiguous.

For more than a decade, Steve's most recent area of interest has been in work on the gravitational self-force.  This refers to the
generally tiny effect the motion of a celestial body experiences due to   
the fact that its movement interacts with its own gravitational field, and it explains why two black holes in orbit around each other will eventually merge.  The focus of Steve's effort was for a future, space-based, gravitational wave detector, originally called LISA.  The perturbative methods which have been used in calculations for decades are ideal for this work.  Steve's valued contribution was in convincing the community of what should actually be compared between different calculations, and in carrying out the computations to high numerical precision.  The self-force world was now waiting for him to take this work to the next level and produce second order results to compare with post-Newtonian theory.

One of Steve's very creative ideas, which is barely known among relativists, but is quite well known in the pulsar timing community, was his suggestion to use precise pulsar timing to detect gravitational waves from the collision of supermassive black holes in the centers of other galaxies.  These very large wavelength waves would cause fluctuations in the arrival times of signals from pulsars --- rotating neutron stars in our own galaxy --- and the accumulation of the impact of these effects from all over the universe would represent an inescapable, noise-like, signal in the timing precision.  This was a brilliant idea, far ahead of its time, leading to a technique that is only now coming into fruition. Pulsars were not even known when Einstein developed his general relativity theory. Though he might have been skeptical about the success of Steve's suggestion, he would certainly have appreciated its inspirational character.  And when it does work, Steve will be fully vindicated.

Among his personal traits, Steve was perhaps most admired for his fearless attitude and his fertile creativity. It may be strange to speak of a creative mind when describing a scientist. After all, science is supposed to be precise and exact, so where can creativity fit in?  But the best scientists discover fresh insights into the workings of the world, and this does indeed require a creative mind. Steve's mind was full of insights, and we have all benefited greatly from his unique ways of thinking.  This may explain why Steve was admired, but it is not why he was loved by anyone who had the chance to know him and spend time with him. It is not just that he would come across as a friendly, nice, and decent guy.  In addition, Steve had a heightened empathy that allowed him to establish a deep and lasting rapport with new friends and colleagues, often almost immediately.

Steve's scientific contributions span four decades, with many   
breakthrough works and deeply original ideas substantiating the impact he  has had on the field of general relativity.  Yet he was nominated for a Fellowship of the American Physical Society only in 2013. This was long overdue. His citation was ``for his many and varied contributions to gravitational physics, which include the computation of black-hole quasi-normal modes, the elucidation of pulsar timing to measure gravitational waves, and foundational contributions to the gravitational self-force.''  We should also note Steve's prescient work on Kaluza-Klein theory, which appeared in his most cited paper and yet Steve barely made any reference to it.  Suffice it to say that Steve had diverse interests, and colleagues were still referring to him for input.  He is already sorely missed.

\vfill\eject

\section*{\centerline
{Remembering Sergio Dain}}
\addcontentsline{toc}{subsubsection}{
\it  Remembering Sergio Dain, by Luis Lehner and Manuel Tiglio}
\parskip=3pt
\begin{center}
Luis Lehner, Perimeter Institute
\htmladdnormallink{llehner-at-perimeterinstitute.ca}
{mailto:llehner@perimeterinstitute.ca}
\end{center}
\begin{center}
Manuel Tiglio, University of California San Diego
\htmladdnormallink{tiglio-at-ucsd.edu}
{mailto:tiglio@ucsd.edu}
\end{center}

It is with our deepest sadness that we report on the sudden
passing away of our friend and colleague Professor Sergio Dain.
Sergio lost his brave yet short battle to cancer on the 24th of February 2016, at
an early age of 46.

Sergio got his Licenciatura and PhD at the University of C\'ordoba in Argentina in
1993 and 1999 respectively, spending a significant fraction of his PhD studies at the
Albert Einstein Institute (AEI) for Gravitational Physics in Golm, Gemany.
The research  themes for his Licenciatura and PhD were on asymptotically flat spacetimes
and topics in gravitational radiation. Afterwards, he spent several years at the
AEI as a postdoctoral researcher, before returning to C\'ordoba in 2006 as a faculty member
and as an independent investigator at CONICET.
After his PhD, Sergio's interests transitioned to geometrical analysis, where he made important
contributions to the initial data problem in General Relativity and conserved quantities in black hole
collisions. He then produced a series of seminal papers establishing geometrical inequalities
between angular momentum and mass in General Relativity. These influential results were recognized
in numerous ways, in particular through invitations to give plenary lectures at, for example,
the 20th International Conference on General Relativity and Gravitation and the 10th Amaldi meeting,
held at Warsaw in 2013, as well as at the Fields Institute at Toronto in 2015.

Sergio had deep and broad interests, always generously shared his research ideas and was a
great mentor and role model. About a dozen undergraduate and graduate students received their
degrees in Argentina under his supervision, and he also mentored four post-docs.
Beyond his research activities, Sergio was avidly engaged in the promotion of
Science in Argentina at different levels, as well as to the General Relativity
community serving, for instance, on the Editorial Board of General Relativity and Gravitation.

Beyond all of the above, to many of us, Sergio was a dear, exceptional friend and an
outstanding person. We will always remember him for his sensitivity, fondness for the arts,
his great sense of humor, contagious laughter, unconditional friendship, and a deep devotion to science.

\vfill\eject

\section*{\centerline
{Is there a place for gravitational physics} 
\centerline
{in the modern, corporate university?}}
\addtocontents{toc}{\protect\medskip}
\addtocontents{toc}{\bf Editorial:}
\addcontentsline{toc}{subsubsection}{
\it Gravitational physics in the modern university, by David Garfinkle}
\parskip=3pt
\begin{center}
David Garfinkle
\htmladdnormallink{garfinkl-at-oakland.edu}
{mailto:garfinkl@oakland.edu}
\end{center}

\vskip0.5truein
(Note: the opinions expressed in this editorial are solely those of the author and do not represent any sort of official pronouncement by either DGRAV or APS).
\vskip0.25truein
Over the past two decades or so, some disturbing trends have taken place in universities in the US: tuition has gone way up (as has student debt). State support for public universities has gone way down.  The percentage of faculty who are tenured or tenure track keeps going down.  The number of administrators keeps going up.  And more and more university boards are taking the attitude that universities should be run like businesses.  This is bad news for everyone.  However, I will argue that it is even worse news for physics than for other areas of academic endeavor, and even worse news for gravitational physics than for other areas of physics. 

First let's consider what is driving these trends.  The current trend in state level politics is for state governments to spend less money.  Since broad support for intellectual pursuits has always been shaky at best in the US, when state governments think about what to cut, support for higher education is always at or near the top of the list. In response, state universities make up for lost revenue by raising tuition, and cut costs by hiring more part time faculty and fewer full time faculty.  In principle, these trends in state universities need not have affected private universities at all.  However, private universities, especially elite ones, have generally positioned their tuition higher than that of state universities and have marketed themselves as using that higher tuition to provide a higher quality education.  Thus, when state universities raised tuition, it is not surprising that private universities followed suit.  

However, there is one point missing from the previous analysis: if universities were really motivated to cut costs, wouldn't they hire fewer administrators, not more?  What then explains the explosion of ``administrative bloat''?  One possible answer is provided by Benjamin Ginsberg in his book {\it The Fall of the Faculty}.  Ginsberg argues that administrative bloat is driven by the need of administrators to feel important.  Thus, for example, the more people who are working for the Dean, the more important he feels, and since administrative tasks don't actually have to be productive, it is easy for the Dean to invent activities for the administrators in his office to do. 

If Ginsberg's analysis is correct, then from an administrator's point of view, both students and faculty are merely means to generate revenue that can be used to hire more administrators.  Thus the administration of a university would be motivated to enroll as many students as possible and to charge them as high a tuition as possible; and to hire as few faculty as possible, with the largest teaching load possible, and at the lowest salary possible.  That is, new faculty hires are likely to be overwhelmingly non-tenure track faculty with such a high teaching load that they have no time at all for scholarship.  

However, the administration cannot hire all part time, non-tenure track faculty, because some tenure track faculty are needed to teach the upper level courses in each department taken by majors and graduate students.  Thus the administration of a university is likely to target those few tenure track hires to the departments with the largest number of majors.  This is bad news for physics, since the number of physics majors tends to be rather small compared to those in other departments.  

With a small number of majors, a physics department in a modern university is likely to be regarded by the administration as a service department, delivering its credit hours in introductory courses taken by those students whose major requires an introductory physics course, such as engineering or health science.  Thus, if the administration hires any tenure track physicists at all, they are likely to be in those areas that most relate to those departments served by the physics department: areas such as materials science or biophysics, but certainly not gravitational physics.  Thus the rise of the modern, corporate, administrative university is bad news for scholarship in all areas.  But it is even worse news for physics than for other departments, and even worse news for gravitational physics than for other areas of physics. 

Of course, there will always be elite universities who justify their high tuition in part by the prestige of their faculty, and these universities will continue to hire tenure track physicists (possibly including gravitational physicists) who will continue to do excellent research.  And from the point of view of the administrations of these elite universities, it is an additional motivating factor that the physicists they hire can bring in grant money whose overhead can be used to hire more administrators.  However, these bright spots, important as they are, are likely to become ever fewer and farther between if the trend of the modern, corporate, administrative university continues.  

What (if anything) can be done about the problem of administrative bloat in the modern, corporate university?  Actually, it is somewhat surprising that steps have not already been taken to curtail this problem.  Recall that the money wasted on hiring superfluous administrators comes from students, their families, state governments, and the federal government.  Together this could be a powerful constituency that should be highly motivated to do something about the problem.  However, the modern, corporate university makes use of an extensive modern public relations apparatus to construct and manipulate its image.  Rather than do something about administrative bloat, this PR apparatus simply denies that the problem exists and blames rising tuition and rising number of administrators on other factors, such as smaller state support, rising government regulation, rising costs of employee health care, and supposed inefficiency of faculty practices.  This last claim is especially dangerous, as it also provides an excuse for university policies that take ever more decisions out of the hands of faculty and give them to administrators.  Both individual faculty and the AAUP (American Association of University Professors) have long sounded the alarm about administrative bloat and about the increasing tendency to treat running a university as though it were the same as running a business.  However, against the professional PR apparatus of the university boards and administrations, the AAUP and individual faculty are losing the war of words.  Perhaps it is time for professional organizations (such as APS) to weigh in on this issue.     

\end{document}